\documentclass[preprint,12pt,authoryear]{elsarticle}
\nopreprintlinetrue

\usepackage[margin=1in]{geometry}
\usepackage{graphicx}
\usepackage{pdflscape}
\usepackage[hidelinks]{hyperref}
\usepackage{xcolor,soul}
\usepackage{longtable}
\usepackage{booktabs}
\usepackage{caption}
\usepackage{subcaption}
\usepackage{multirow}
\usepackage{natbib}
\usepackage{amsmath}

\usepackage{titlesec}
\titleformat{\section}
{\normalfont\large\bfseries}{\thesection}{1em}{}
\titleformat{\subsection}
{\normalfont\normalsize\bfseries}{\thesubsection}{1em}{}
\titleformat{\subsubsection}
{\normalfont\normalsize\bfseries}{\thesubsubsection}{1em}{}

\begin{document}

\begin{frontmatter}

\title{Semi-on-Demand Off-Peak Transit Services with Shared Autonomous Vehicles --- 
Service Planning, Simulation, and Analysis in Munich, Germany}

\author[nwu]{Max T.M. Ng}
\author[tum]{Roman Engelhardt}
\author[tum]{Florian Dandl}
\author[nwu]{Vasileios Volakakis}
\author[nwu]{Hani S. Mahmassani\corref{cor1}}
\author[tum]{Klaus Bogenberger}

\cortext[cor1]{Corresponding Author}

\affiliation[nwu]{Northwestern University Transportation Center, 600 Foster Street, Evanston, IL 60208, USA}
\affiliation[tum]{Chair of Traffic Engineering and Control, Technical University of Munich, Arcisstraße 21, 80333 Munich, Germany}

\begin{abstract}
This study investigates the implementation of semi-on-demand (SoD) hybrid-route services using Shared Autonomous Vehicles (SAVs) on existing transit lines.
SoD services combine the cost efficiency of fixed-route buses with the flexibility of on-demand services.
SAVs first serve all scheduled fixed-route stops, then drop off and pick up passengers in the pre-determined flexible-route portion, and return to the fixed route.
This study addresses four key questions: optimal fleet and vehicle sizes for peak-hour fixed-route services with SAVs and during transition (from drivers to autonomous vehicles), optimal off-peak SoD service planning, and suitable use cases.
The methodology combines analytical modeling for service planning with agent-based simulation for operational analysis.
We examine ten bus routes in Munich, Germany, considering full SAV and transition scenarios with varying proportions of drivers.
Our findings demonstrate that the lower operating costs of SAVs improve service quality through increased frequency and smaller vehicles, even in transition scenarios.
The reduced headway lowers waiting time and also favors more flexible-route operation in SoD services.
The optimal SoD settings range from fully flexible to hybrid routes, where higher occupancy from the terminus favors shorter flexible routes.
During the transition phase, limited fleet size and higher headways constrain the benefits of flexible-route operations.
The simulation results corroborate the SoD benefits of door-to-door convenience, attracting more passengers without excessive detours and operator costs at moderate flexible-route lengths, and validate the analytical model.
\end{abstract}


\end{frontmatter}

\section{Introduction}

\subsection{Motivation}
The advent of shared autonomous vehicles (SAVs) brings new perspectives to enhancing multimodal public transit services, particularly as feeders in less dense areas \citep{ng_redesigning_2024}. 
With lower operating costs, they can address the first-mile-last-mile problem and bring passengers to other transit modes more efficiently than conventional ride-sharing. 
While transit agencies retain some drivers due to union contracts or operational reasons, 
the lower operating costs of SAVs also allow for a larger fleet of smaller vehicles and thereby more frequent services \citep{hatzenbuhler_transitioning_2020}. 
Transit agencies can plan their fleet (vehicle size and type) according to peak-hour fixed-route service, which leaves redundant capacity at off-peak times for more flexible services and expand their catchment areas. 
For example, smaller but more SAVs can pick up and drop off some passengers at their homes, bringing more convenience and stimulating more demand \citep{frei_flexing_2017}.

\begin{figure}[thpb]
	\center
	\includegraphics[width=\textwidth]{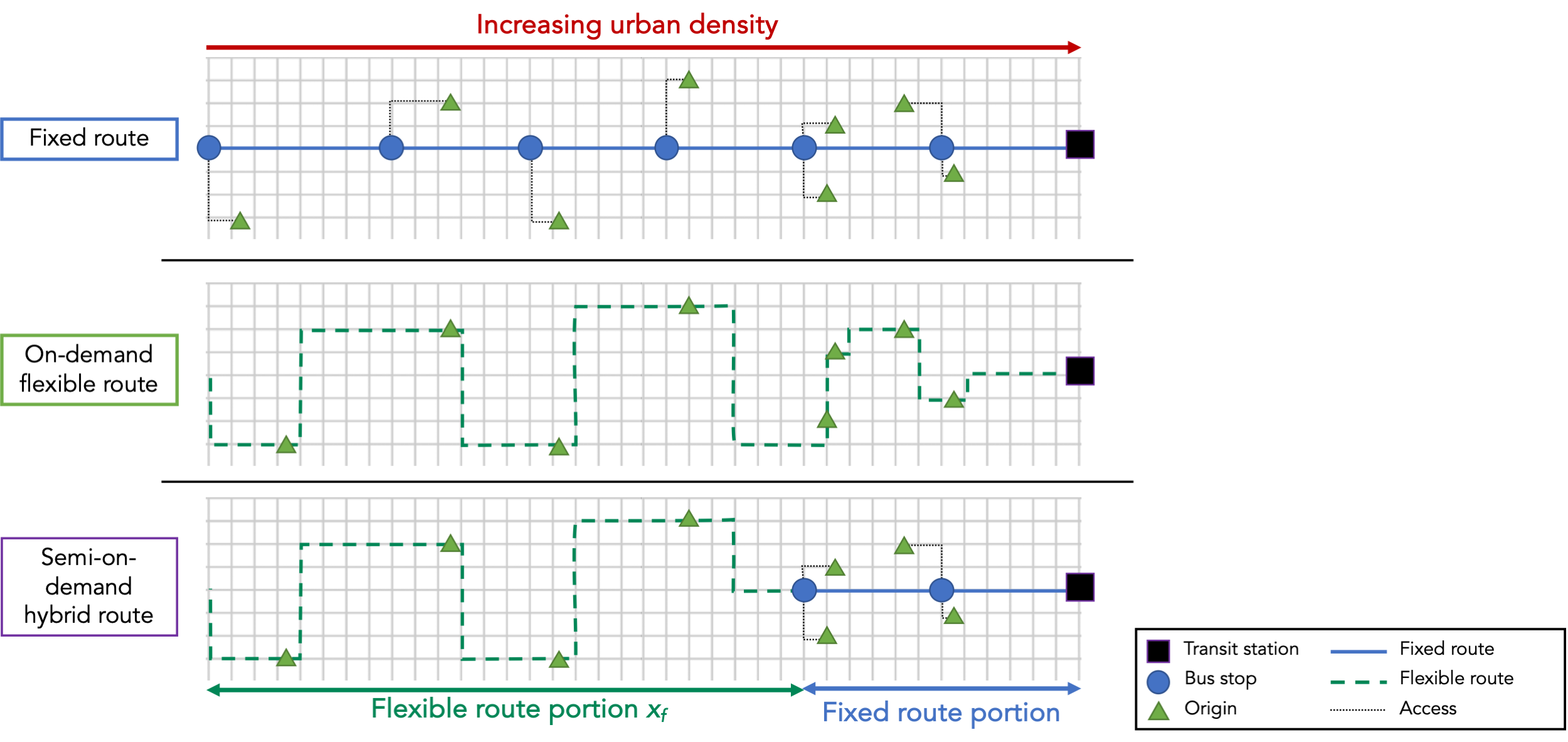}
	\caption{Illustration of semi-on-demand hybrid route as a feeder service}
        \label{fig:route_forms}
\end{figure}

In semi-on-demand (SoD) hybrid-route services investigated in previous studies \citep{ng_autonomous_2023, ng_semi--demand_2024} (Figure~\ref{fig:route_forms}),
SAVs first serve all fixed-route stops based on a schedule, then drop off and pick up passengers in the flexible-route portion (pre-determined with length $x_f$), and return to fixed-route scheduled stops and the terminus.
It is shown to combine the cost efficiency of fixed-route buses in denser areas with the flexibility of on-demand services in less populated regions.

\subsection{Problem Description}

This study considers improving current transit services with SAVs by enhancing peak-hour fixed-route service (in the existing format but more frequent due to SAV cost savings) and implementing off-peak SoD hybrid-route services in the same route catchment area. 
We address four primary research questions:
\begin{enumerate}
    \item At peak hours, what are the optimal fleet and vehicle sizes for current-day transit lines with the lower operating costs of SAVs, when we minimize waiting time and operator's cost subject to capacity and budget constraints?
    \footnote{Determining fleet and vehicle size with fixed routes at peak hours ensures sufficient capacity to satisfy demand. This also leads to more conservative estimates of the benefits of using smaller vehicles to provide flexible services at off-peak.}
    \item During a transition phase from drivers to full automation, how would the operating budget affect the previous optimal results?
    \item At off-peak hours, what is the optimal service planning (schedule and route form) for SoD hybrid-route services, considering stochastic requests and balance between schedule time, request rejection, and operational performance?
    \item What are the use cases of off-peak SoD services considering benefits and impacts to passengers and the operator? 
\end{enumerate}

For SoD, too long flexible routes would lead to excessive detours, longer journey times, and higher operating costs. 
Meanwhile, smaller headways (e.g., more smaller vehicles) favor longer flexible routes by reducing the detours of each trip \citep{ng_semi--demand_2024}.
Therefore, optimizing SAV fleet and vehicle sizes for peak-hour fixed routes may also benefit off-peak SoD hybrid-route services.

The problem scope is defined with some key assumptions: 
(i) continuous operations during the study hours; 
(ii) inelastic demand (yet sensitive to walking distance); 
(iii) requests made when passengers are ready for boarding, with no reservations considered; and
(iv) individual travelers, i.e. no group bookings are considered.

\subsection{Contributions}

Building upon the theoretical formulations of costs and benefits derived in previous work \citep{ng_semi--demand_2024}, this study focuses on the service planning, simulation, and analysis of the SoD hybrid-route transit feeders using SAVs. 
It presents conceptual, theoretical, and methodological contributions:

\begin{enumerate}
    \item We formulate the service planning problem of peak-hour fixed-route services in terms of optimal fleet size and vehicle size, considering changes in costs and retention of drivers. The study further conceptualizes the off-peak SoD hybrid-route service.
    \item We derive analytical results of convex optimization for the peak-hour fleet and vehicle sizes (in full-SAV and partial-driver scenarios) and off-peak schedules considering detours, service guarantee, and peak/off-peak service level. This is a timely addition to the literature for planning the gradual introduction of SAVs in transit service.
    \item By analyzing the agent-based simulation results on ten existing bus routes in the city of Munich, Germany, we optimize the flexible-route portion $x_f$ and headway $h$, and examine the use cases and benefits of hybrid routes (in terms of access, waiting, and riding times for users and vehicle distances and requirements for operators). We contrast theoretical predictions with simulation results and study variations in user experience.
\end{enumerate}

\begin{figure}[thpb]
	\center
	\includegraphics[width=\textwidth]{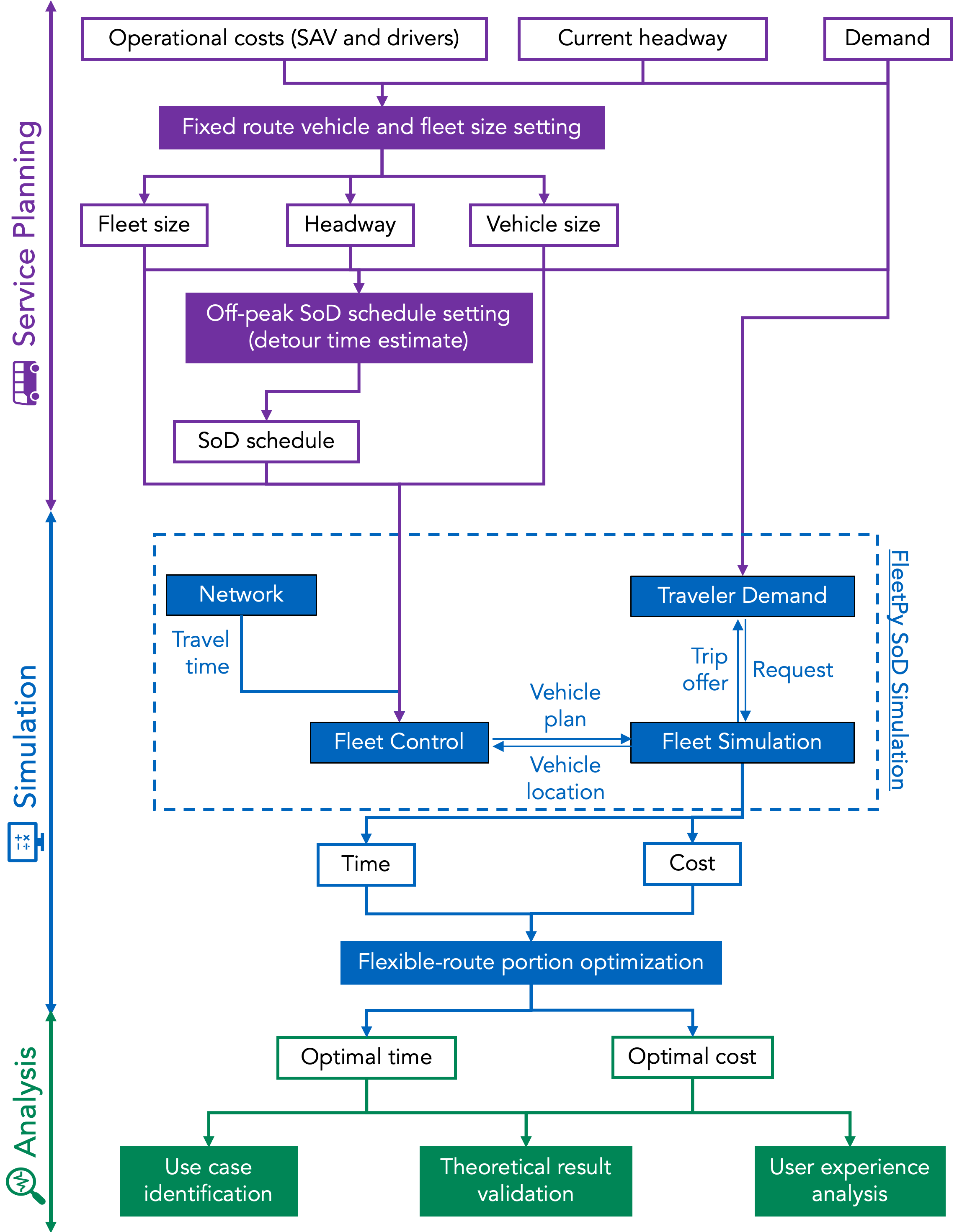}
	\caption{Flowchart of the service planning, simulation, and analysis}
        \label{fig:flowchart}
\end{figure}

Figure~\ref{fig:flowchart} summarizes the flow of the methodology. 
The remainder of the paper is structured as follows.
Section~\ref{sec:bg} presents an overview of relevant literature on demand-responsive transit, SAVs in public transit, and ride-sharing simulation.
Section~\ref{sec:method} follows with the methodology, illustrated with an experiment described in Section~\ref{sec:exp} and results presented in Section~\ref{sec:res}.
Section~\ref{sec:concl} concludes the paper.

\section{Background}
\label{sec:bg}

\subsection{Demand-Responsive Transit}

Research on demand-responsive transit (DRT) spans a few decades (see surveys by \cite{errico_survey_2013, vansteenwegen_survey_2022}). 
Several studies have explored flexible transit route designs, suggesting conditions to use demand-responsive over fixed-route services \citep{li_feeder_2010} and cost advantages for low-to-moderate passenger demand \citep{nourbakhsh_structured_2012}.
Recent research has focused on integrating DRT with traditional fixed services, 
e.g., supplementary DRT improvements to fixed-route bus services \citep{chen_analysis_2017, fielbaum_improving_2024} and 
combination of fixed-schedule and DRT services during off-peak hours and in the suburbs \citep{calabro_adaptive_2023}.

Various optimization strategies have been employed in semi-flexible transit system research, including analytical models \citep{qiu_dynamic_2014}, metaheuristics \citep{galarza_montenegro_large_2021}, stochastic approaches with Markov decision processes \citep{rambha_adaptive_2016, li_frequency-based_2023}, and agent-based simulation \citep{rich_fixed_2023}. 

However, most studies focus on the performance of flexible-route services or the decision to deploy either flexible-route or fixed-route services. 
In contrast, this study builds on previous work \citep{ng_semi--demand_2024} to design and simulate hybrid-route service, representing a continuum between these two service modes.

\subsection{Shared Autonomous Vehicles in Public Transit Systems}

SAVs have shown the potential in expanding transit coverage and enhancing service attractiveness \citep{salazar_intermodal_2020, shen_integrating_2018}
(also see a survey \citep{narayanan_shared_2020}).
Several studies have explored the synergistic relationship between SAVs and public transit systems \citep{shen_integrating_2018, cortina_fostering_2023, gurumurthy_first-mile-last-mile_2020}, sometimes by integrating SAVs into traditional transit \citep{levin_linear_2019}. 
Conversely, some considered SAVs as competitors \citep{sieber_improved_2020}, with performance improvements by replacing fixed bus routes with flexible routes in suburbs in Chicago, USA \citep{ng_autonomous_2023, volakakis_city-wide_2023}.
The joint design of transit and SAVs has been another area of focus, e.g., optimizing vehicle size \citep{alonso-mora_-demand_2017}, fleet size \citep{pinto_joint_2020, dandl_regulating_2021}, network \citep{ng_redesigning_2024}, and timetabling and vehicle scheduling \citep{cao_autonomous_2019}.

Automation can also benefit line-based buses, with a general shift towards higher frequencies and smaller vehicles \citep{bosch_cost-based_2018} found to attract passengers through improved service \citep{hatzenbuhler_transitioning_2020} and SAVs as feeders to improve efficiency \citep{badia_design_2021}.

In summary, the integration of SAVs in public transit systems presents opportunities for improved service quality, expanded coverage, and enhanced operational efficiency. 
However, challenges remain in optimizing the balance between traditional fixed-route services and more flexible autonomous options across urban contexts and demand patterns.

\subsection{Ride-Sharing Simulation}
Agent-based simulation has been widely used to represent the dynamic and stochastic nature of underlying vehicle routing problems and the interaction between operators and riders.
Studies focused on solving the online assignment problem \citep{alonso-mora_-demand_2017, Simonetto.2019, Wang.2023}
or repositioning idle vehicles to accommodate future expected demand \citep{Zhang.2016, Tafreshian.2021, Engelhardt.2023}.
Apart from door-to-door services, short walking legs for riders \citep{Engelhardt.2021, Fielbaum.2021} or transfers between vehicles \citep{Masoud.2017,Namdarpour.2024} have been proposed to increase efficiency and sharing rate for the service.

On the operational aspects of demand-responsive services,
models like SUMO \citep{armellini_tool_2021}, MATSim \citep{horni_multi-agent_2016}, and Polaris \citep{gurumurthy_integrating_2020, cokyasar_dynamic_2022} have been extended to simulate both fixed- and flexible-route services. 
Other studies have developed new agent-based simulation models focusing on operational aspects or computational performance \citep{alonso-mora_-demand_2017, fagnant_dynamic_2018, dandl_evaluating_2019, engelhardt_fleetpy_2022}.
These studies focused on the supply side, 
i.e., how to serve a fixed set of riders as efficiently as possible.
Meanwhile, agent-based demand models \citep{Ruch.2018, Gurumurthy.2020, Wilkes.2021} have been developed to estimate the demand attracted to the new modes.

This study adapts the open-source Python package Fleetpy \citep{engelhardt_fleetpy_2022} to model the hybrid-route SoD service and the interaction between riders, operator, and vehicles.

\section{Methodology}
\label{sec:method}

This section first discusses the service planning for peak-hour fixed-route service in full SAV and transition scenarios,
followed by off-peak SoD schedule setting.
It then describes the simulation of SoD service and result analysis.

The notations used are summarized in Table~\ref{tab:notation}.
\footnote{Sets are denoted as capital scripted characters, constants as Greek or capital Roman characters, random variables (R.V.) as capital Roman characters, and indices and variables as small Roman characters. Superscripts are qualifiers and subscripts are indices.}

\begin{longtable}{@{}p{1.2cm}p{10cm}p{4cm}@{}} 
\caption{Notation}
\label{tab:notation} \\
\toprule
Symbol & Description & Experiment Value \\
\midrule
\endfirsthead

\multicolumn{3}{c}%
{{\bfseries \tablename\ \thetable{} -- continued from previous page}} \\
\toprule
Symbol & Description & Experiment Value \\
\midrule
\endhead


\bottomrule
\endlastfoot

$a_{i,z}$ & Arrival time of passenger $i$ according to plan $z$ & \\
$b \in \mathcal{B}$ & Vehicle size & $\mathcal{B} = \{5,8,20,44,70\}~pax$\\
$c^g$ & Total generalized cost & \\
$c^o_v$ & Vehicle cost of vehicle $v$ & \\
$c^u_r$ & User cost of rider $r$ & \\
$C_v$ & Vehicle capacity of vehicle $v$ & \\
$c(b,h)$ & Cost function of a route with vehicle size $b$ and headway $h$ & \\
$d_r$ & Destination node of request $r$ & \\
$d_r^f$ & Nearest bus stop to the destination of request $r$ & \\
$d_{v,z}$ & Distance traveled by vehicle $v$ according to plan $z$ & \\
$D$ & Total detour distance (R.V.) & \\
$D^p$ & Peak demand & \\
$D^w$ & Maximum walking distance & 500~m \\
$E_i$ & Earliest start time for stop $i$ & \\
$\mathcal{E}$ & Edges of the network & \\
$\mathcal{G}$ & Street network & \\
$h$ & Headway & \\
$h^*$ & Optimal headway & \\
$h^p$ & Peak-hour headway & \\
$H^0$ & Existing headway & \\
$i \in \mathcal{I}$ & Stop & \\
$L^x$ & Route length & \\
$N^a$ & Number of SAVs & \\
$N^{d,0}$ & Current number of drivers & \\
$n^r_z$ & Number of requests satisfied in vehicle plan $z$ & \\
$n^s_z$ & Number of requests served in the fixed route in vehicle plan $z$ & \\
$N$ & Number of flexible-route requests (R.V.) & \\
$\mathcal{N}$ & Nodes of the network & \\
$o_r$ & Origin node of request $r$ & \\
$o_r^f$ & Nearest bus stop to the origin of request $r$ & \\
$r \in \mathcal{R}$ & Customer request  & \\
$s$ & Fleet size & \\
$t^a_r$ & Access time of request $r$& \\
$t^d_c$ & Allowable vehicle detour time & \\
$t^t_r$ & Riding time of request $r$& \\
$t^v_v$ & Time deployed for vehicle $v$ & \\
$t^w_r$ & Waiting time of request $r$ & \\
$t_r$ & Request time of request $r$ & \\
$T$ & Total expected detour time (R.V.) & \\
$T^c$ & Cycle time & \\
$T^l_i$ & Latest start time for stop $i$ & \\
$T^s$ & Dwell time & 30~s \\
$T^s_i$ & Duration of stop $i$ & \\
$T^{w, max}$ & Maximum waiting time & 15~min \\
$v \in \mathcal{V}$ & Vehicle & \\
$V^d$ & Constant vehicle speed (in planning) & 40~km/h \\
$X$ & Irwin-Hall distribution R.V. & \\
$Y$ & Detours for requests (R.V.) & \\
$Y_i$ & Stop location of stop $i$ & \\
$z$ & SoD schedule & \\
$\alpha$ & Portion of original drivers remaining in the workforce & \\
$\gamma^a$ & Cost coefficient for access time & 2 \\
$\gamma^c_b$ & Operational costs (operating + capital) of SAV size $b$ & \{4.4, 5.9, 11.05, \\ & & 16.2, 23.8, 33.9\} (€/h) \\
$\gamma^{c,0}$ & Current bus HDV operational cost & €36.3/h \\
$\gamma^d$ & Driver cost & €15.3/h \\
$\gamma^f$ & Reward coefficient for fixed-route requests & 1e+6 \\
$\gamma^m_b$ & Operating costs of SAV size $b$ per hour & \{2.1, 2.6, 4.15, \\ & & 5.7, 9.5, 12.6\} (€/h) \\
$\gamma^{m,0}$ & Current bus HDV operating cost & €24.8/h \\
$\gamma^o_v$ & Cost coefficient for vehicle distance of vehicle $v$ & \\
$\gamma^r$ & Cost coefficient for traveler time & €16.5/h \\
$\gamma^s$ & Reward coefficient for requests satisfied & 1e+6 \\
$\gamma^t$ & Value of time & €16.5/h \\
$\gamma^v_v$ & Cost coefficient for vehicle time of vehicle $v$ & \\
$\gamma^w$ & Cost coefficient for waiting time & 1.5 \\
$\lambda$ & Poisson distribution parameter & \\
$\Lambda$ & Demand & \\
$\phi^p$ & Maximum in-vehicle travel time factor & 2 \\
$\rho$ & Capacity buffer & 0.9 \\
$\phi(z)$ & Objective function for schedule $z$ & \\
\end{longtable}

\subsection{Service Planning}

\subsubsection{Peak-Hour Fixed-Route Service --- Optimal Vehicle and Fleet Size (Full SAV Scenario)}

For a peak-hour fixed route with 100\% SAVs, we develop a convex optimization problem with analytical closed-form solutions of vehicle size $b$ and headway $h$, subject to capacity and budget constraints.
The goal is to minimize $c(b,h)$ in Eq.~\eqref{eq:C_b_h}, the weighted sum of users' waiting time and operator's cost (including both capital and operating expenses).
The first term is the waiting cost with the value of time $\gamma^t$, waiting cost coefficient $\gamma^w$, and demand $\Lambda$.
The second term is the operational (capital and operating) cost to run a fleet with $s = T^c / h$ vehicles with the unit cost per time $\gamma^c_b$ and cycle time $T^c$.

    \begin{align}
    c(b,h) &=
    \frac{1}{2} \gamma^t \gamma^w \Lambda h
    + \gamma^c_b \frac{T^c}{h}
    \label{eq:C_b_h}
    \end{align}

For the optimality condition with respect to $h$, we set the first partial derivative in Eq.~\eqref{eq:C'_b_h} to zero. 
The cost function is also shown to be convex as the second derivative in Eq.~\eqref{eq:C''_b_h} is always positive.

    \begin{align}
    \frac{\partial c}{\partial h} &=
    \frac{1}{2} \gamma^t \gamma^w \Lambda
    - \gamma^c_b \frac{T^c}{h^2}
    \label{eq:C'_b_h}
    \\
    \frac{\partial^2 c}{\partial h^2} &=
    2 \gamma^c_b \frac{T^c}{h^3} > 0
    \label{eq:C''_b_h}
    \end{align}

This leads to the optimal headway $h^*$ in Eq.~\eqref{eq:h_opt},
resulting in the minimum cost $c^* (b)$ in Eq.~\eqref{eq:C_opt}.
$h^*$ is inversely proportional to the square root of demand density, which bears similarity to the result of \citep{newell_issues_1979}.

         \begin{align}
    h^* &= \sqrt{\frac{2\gamma^c_b T^c}{\gamma^t \gamma^w \Lambda}} 
    \label{eq:h_opt}
    \\
    c^*(b) &= \sqrt{2\gamma^t \gamma^w \gamma^c_b \Lambda T^c }
    \label{eq:C_opt}
    \end{align}

To ensure sufficient line capacity, $h$ is bounded above in Eq.~\eqref{eq:h_UB} (given vehicle size $b$ and capacity buffer $\rho \in [0,1]$) to satisfy peak demand $D^p$.
To find the lowest-cost solution that provides sufficient capacity, we iterate through all discrete $b$.

    \begin{align}
    h &\leq \frac{b \rho}{D^p}
    \label{eq:h_UB}
    \end{align}

To comply with the existing operating budget, $h$ is bounded below in Eq.~\eqref{eq:h_LB} (with existing operating unit cost $\gamma^{c,0}$ and headway $H^0$).
Combining this with Eq.~\eqref{eq:h_UB} brings us Eq.~\eqref{eq:b_LB}, which checks whether a vehicle size can fulfill both capacity and budget constraints.

    \begin{align}
    h &\geq \frac{\gamma_b^c}{\gamma^{c,0}} H^0
    \label{eq:h_LB}
    \\
    b &\geq \frac{\gamma_b^c}{\gamma^{c,0}} \frac{H^0 D^p}{\rho}
    \label{eq:b_LB}
    \end{align}

In short, we find $h^*$ in Eq.~\eqref{eq:h_opt} subject to the upper bound in Eq.~\eqref{eq:h_UB} and lower bound in Eq.~\eqref{eq:h_LB}.
If Eq.~\eqref{eq:b_LB} is violated, the upper bound for $h$ is lower than the lower bound, and there is no feasible solution for that particular vehicle size $b$.

\subsubsection{Peak-Hour Fixed-Route Service --- Optimal Vehicle and Fleet Size (Transition Scenario with Drivers)}
\label{sec:transition}
The transition scenario is motivated by a socially responsible path towards SAV-based public transit that retains drivers in the workforce until their redeployment/retirement.
For simplicity, they are assumed to drive some SAVs in the new fleet (same vehicle size for each route).
We adapt the previous optimization procedure by detaching capital costs from overall operational cost,
under the assumption that the capital investment decision is already made for a long-term full rollout scenario.
Hence, the equations in the previous subsection are still valid, except that $\gamma^c_b$ is replaced by $\gamma^m_b$.
The question is whether we can still improve some services by replacing a bus with smaller SAVs while maintaining budget balance and satisfying capacity.

The operating budget in the transition period is composed of labor costs ($\gamma^d$ per remaining driver) and SAV operating costs (unit cost $\gamma^m_b$).
When a portion $\alpha$ of the original $N^{d,0}$ drivers remain in the workforce,
and an additional SAV fleet of $N^a$ is added to provide services, i.e., the fleet size becomes $s = \alpha N^{d,0} + N^a$.
The total operating costs are subject to the current budget (unit operating cost $\gamma^{m,0}$ and labor cost) in Eq.~\eqref{eq:op_cost^d_constraint}.

    \begin{align}
        \gamma^m_b (\alpha N^{d,0} + N^a) + \gamma^d \alpha N^{d,0} \leq (\gamma^{m,0} + \gamma^d) N^{d,0}
        \label{eq:op_cost^d_constraint}
    \end{align}

This can be rearranged as an upper bound on the fleet size $s$ in Eq.~\eqref{eq:op_i_UB} or a lower bound on the headway in Eq.~\eqref{eq:op_h_LB}.

    \begin{align}
        s &\leq \frac{\gamma^{m,0} - \gamma^d (1-\alpha)}{\gamma^m_b} N^{d,0}
        \label{eq:op_i_UB}
        \\
        h &\geq
        \frac{\gamma^m_b}{(\gamma^{m,0} - \gamma^d (1-\alpha))} H^0
        \label{eq:op_h_LB}
    \end{align}

\subsubsection{Off-Peak Semi-on-Demand Service --- Schedule Setting}

An SoD schedule is formed by fixed stops (existing route schedule) and the flexible-route portion (with length $x_f$ in Figure~\ref{fig:route_forms}, existing journey time plus allowable vehicle detour time $t^d_c$). 
Serving a certain level $c$ (say 95\%) of stochastic demand in the flexible-route portion requires an adequate $t^d_c$. 
We derive its cumulative distribution function with the assumptions of 
(i) independently uniformly distributed detours for each request ($Y \sim \text{Uniform}(0, D^w)$), 
(ii) Poisson occurrences of flexible-route requests ($N \sim \text{Poisson}(\lambda)$), and 
(iii) constant vehicle speed $V^d$.

The total detour distance $D$ given $N$ requests is the sum of detours (multiplied by two for deviating from and returning to the main route) in Eq.~\eqref{eq:D_y}. 
The total detour time given the number of request points, $T|N$, is the sum of dwell times and total detour time in Eq.~\eqref{eq:T|N}.

    \begin{align}
        D|N &= \sum_{r=1}^N 2 Y
        \label{eq:D_y}
        \\
        T|N &= N T^s + \frac{D|N}{V^d} = N T^s + \sum_{r=1}^N \frac{2 Y}{V^d}
        \label{eq:T|N}
    \end{align}

We note that $T|N$, as a sum of uniform distributions ($Y$), follows the Irwin-Hall distribution with a cumulative distribution function (CDF) in Eq.~\eqref{eq:Irwin} \citep{johnson_continuous_1995}.
By considering $\sum_{r=1}^N Y = V^d / 2 (T|N - N T^s)$ from Eq.~\eqref{eq:T|N} and normalizing it by setting $X = \sum_{r=1}^N Y / D^w $ (so that the support of $X|N$ is $[0,N]$),
we obtain the CDF of $T|N$ in Eq.~\eqref{eq:T_n}.

    \begin{align}
        P(X \leq x) &=
        \frac{1}{n!} \sum_{k=0}^{\lfloor x \rfloor} (-1)^k \binom{n}{k} (x-k)^n
        \label{eq:Irwin}
        \\
        P(T \leq t^d_c|N=n) &= P \left( X \leq \frac{V^d}{2 D^w} (t^d_c - n T^s) \right)
        \label{eq:T_n}
    \end{align}

As the occurrence of flexible-route requests $N$ follows a Poisson distribution, its probability mass function is given in Eq.~\eqref{eq:P_N}. 

    \begin{align}
        P(N=n) &= \frac{e^{-\lambda} \lambda^n}{n!}
        \label{eq:P_N}
    \end{align}

By the law of total probability, we get $P(T \leq t^d_c) = \sum_{n=0}^\infty P(T \leq t^d_c|N=n) P(N=n)$. 
This leads to the CDF of $T$ in Eq.~\eqref{eq:T_t^d_c}.
The corresponding $t^d_c$ for the required confidence level of demand satisfaction can then be approximated from the CDF curve.

    \begin{align}
        P(T \leq t^d_c) &= \sum_{n=0}^\infty \frac{e^{-\lambda} \lambda^n}{(n!)^2} \sum_{k=0}^{\lfloor x \rfloor} (-1)^k \binom{n}{k} (x-k)^n, x=\frac{V^d}{2 D^w} (t^d_c - n T^s)
        \label{eq:T_t^d_c}
    \end{align}

Afterward, we consider the constraint posed by the vehicle requirements of a schedule.
Taking fleet size $s$ that is set in the previous subsection based on the peak-hour fixed route operating with a headway $h^p$ and a cycle time of $T^c$,
we have $s=T^c/h^p$.
For SoD services at off-peak hours with the same fleet size $s$ but headway $h < h^p$, 
the additional detour $t^d_c$ poses a more stringent requirement on $s$, i.e., $s \geq (T^c + t^d_c) / h$.
Combining the two equations suggests an upper bound of $t^d_c$ in Eq.~\eqref{eq:t^d_c_bound}.

    \begin{align}
        t^d_c &\leq
        T^c \left( \frac{h}{h^p} - 1 \right)
        \label{eq:t^d_c_bound}
    \end{align}

These equations are used to design schedules for SoD services in different lines and yield statistical insights into the trade-offs between service guarantee, performance, and operating costs.

\subsection{Simulation of Semi-on-Demand Service}
We expand the FleetPy simulation framework \citep{engelhardt_fleetpy_2022} with a scheduled SoD module (Figure~\ref{fig:flowchart}).
It models the interaction among three agents: riders $r \in \mathcal{R}$,
a service operator, and vehicles $v \in \mathcal{V}$ within a street network $\mathcal{G} = (\mathcal{N},\mathcal{E})$ with nodes $\mathcal{N}$ and edges $\mathcal{E}$ in discrete time steps.
Customers request trips (with demand as a function of access distance),
where each rider request $r$ is defined by the request time $t_r$, origin $o_r \in \mathcal{N}$, and destination $d_r \in \mathcal{N}$.
The operator evaluates requests, considers service quality and operational constraints (e.g., maximum waiting and travel times and vehicle capacity),
and assigns requests to individual vehicles operating on SoD schedules $z$.
A schedule is an ordered list of stops that are processed by the vehicle one after another,
whereas a stop $i \in \mathcal{I}$ is characterized by a location $Y_i \in \mathcal{N}$, the latest start time $T^l_i$, the earliest start time $E_i$, a duration $T^s_i$, and sets of requests boarding and alighting the vehicles.

In between stops, vehicles travel in the network on the fastest path.
If they arrive at the next stop $i$ before the scheduled time $E_i$, they wait there until $E_i$.
In the next step, they perform the boarding task associated with this stop for a duration of $T^s_i$ and then go to the next stop.

Requests are assigned by sequential insertion into currently assigned vehicle schedules.
The feasibility of each resulting schedule is then evaluated based on the following constraints: 
(i) each stop $i$ in the schedule is served no later than $T^l_i$; 
(ii) each rider waits less than the maximum waiting time $T^{w, max}$; 
(iii) the maximum in-vehicle travel time of each passenger is within the direct travel time by a factor $\phi^p$; and 
(iv) the number of passengers never exceeds the vehicle capacity $C_v$.

The objective $\phi(z)$ of all feasible schedules is then calculated in Eq.~\eqref{eq:rho} that balances vehicle distance,
rider travel time, and requests served in the fixed-route portion.
 $\gamma^o_v$, $\gamma^t$, $\gamma^r$, and $\gamma^s$ are the cost/reward coefficients for vehicle distance, traveler time, requests satisfied, and requests served in a fixed route, respectively.
 \footnote{$\gamma^s$ prioritizes the assignment of riders to fixed stops if possible.}
 $d_{v,z}$, $a_{r,z}$, $n^{r}_{z}$, and $n^s_{z}$ are respectively the distance traveled by vehicle $v$, arrival time of rider $r$, number of requests satisfied, and number of requests served in a fixed route.

    \begin{align}
         \phi(z)
         &= \sum_{v \in \mathcal{V}} \gamma^o_v d_{v,z}
         + \sum_{r \in \mathcal{R}} \gamma^{t} (a_{r,z}-t_r)
         - \gamma^{r} n^r_{z}
         - \gamma^{s} n^s_{z}
         \label{eq:rho}
    \end{align}
    
The feasible schedule with the minimum objective is then selected and assigned.
The process is summarized as follows:
\begin{enumerate}
    \item If a rider request $r$ is picked up or dropped off in the fixed-route portion, $o_r$ or $d_r$ is shifted to $o_r^f$ or $d_r^f$ corresponding to the nearest bus stop. 
    \item A request is assigned to the vehicle schedule by: 
    (i) adding the request to the boarding or alighting list of a stop if the corresponding locations match; or 
    (ii) inserting a new stop to pick up or drop off the request.
    All feasible insertions for a request and each vehicle are computed by an exhaustive search, and the schedule $z$ that minimizes $\phi(z)$ in Eq.~\eqref{eq:rho} is assigned.
    If no feasible insertion is found, the rider's request is rejected.
    \item Steps 1 and 2 are repeated until all current requests are processed.
    Then, vehicles move and perform boarding processes according to their assigned schedules.
\end{enumerate}

The simulation records user and vehicle data on trip level and at the end of the simulation, metrics in aggregate terms are computed.

\subsection{Analysis}

The simulation results of each scenario (each flexible-route length $x_f$) are analyzed and compared to identify the suitable use cases for SoD.
Generalized costs are used as metrics that include both user's and operator's costs.

The user's cost $c^u_r$ of rider $r$ in Eq.~\eqref{eq:c^u_i} is composed of access (walking), waiting, and riding costs, where $t^a_r$, $t^w_r$, and $t^t_r$ are the access, waiting, and riding times and $\gamma^a$, $\gamma^w$, and $\gamma^r$ are the respective cost coefficients.

    \begin{align}
        c^u_r &= \gamma^a t^a_r + \gamma^w t^w_r + \gamma^r t^t_r
        \label{eq:c^u_i}
    \end{align}
    
The vehicle cost $c_v^o$ for vehicle $v$ in Eq.~\eqref{eq:c^o_v} consists of distance-based operating cost and vehicle marginal cost, where $d_v$ and $t^v_v$ are the distance traveled and time deployed with $\gamma^o$, and $\gamma^v$ as the respective cost coefficients.

    \begin{align}
        c^o_v &= \gamma^o_v d_v + \gamma^v_v t^v_v
        \label{eq:c^o_v}
    \end{align}
    
The total generalized cost $c^g$ in Eq.~\eqref{eq:c^g} is the sum of all users' costs and operator's costs.

    \begin{align}
        c^g &= \sum_{r \in \mathcal{R}} c^u_r + \sum_{v \in \mathcal{V}} c^o_v
        \label{eq:c^g}
    \end{align}

The optimal SoD route form may be fixed route ($x_f = 0$), hybrid route ($0 \leq x_f \leq L^x$), or flexible route ($x_f = L^x$), 
depending on the minimum generalized cost $c^g$ and number of passengers served.
We identify such cases and also analyze the more fine-grained effects on journey time and operational costs,
in comparison with the theoretical values \citep{ng_semi--demand_2024}.

\section{Experiment Design}
\label{sec:exp}

\begin{figure}[thb]
    \center
    \includegraphics[width=0.7 \textwidth]{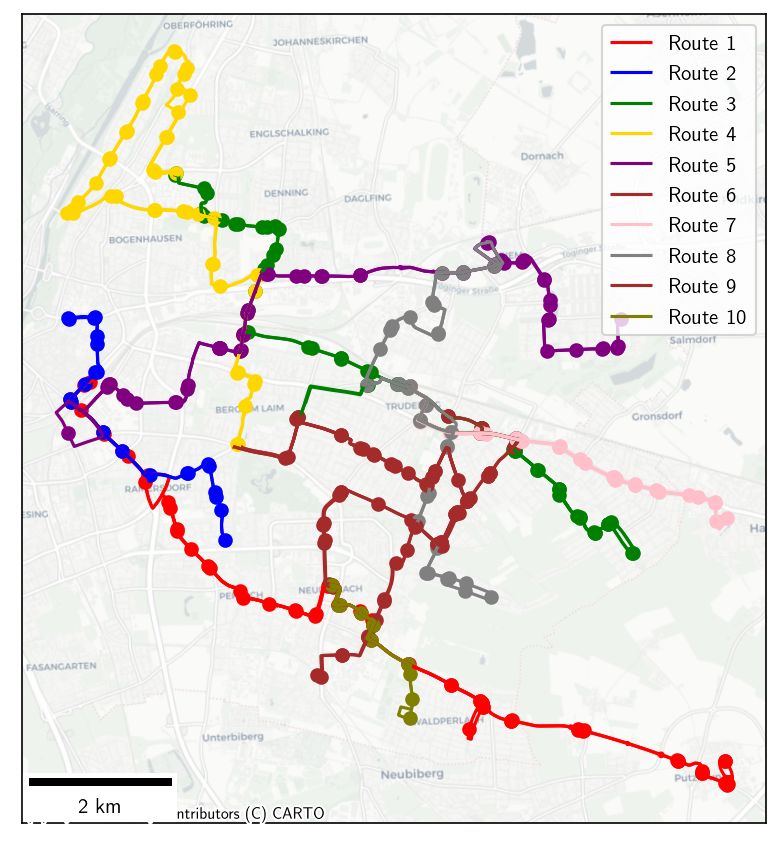}
    \caption{Ten bus routes studied in Munich, Germany}
        \label{fig:exp_buslines}
\end{figure}

We selected ten bus routes in eastern Munich, Germany (Figure~\ref{fig:exp_buslines}) with different characteristics. 
Routes 1, 3, and 5 offer frequent service with at most 10-minute headways, serving as longer routes that connect suburbs to the city center.
Routes 2 and 4 provide less frequent service with 20-minute headways with similar purposes.
The remaining Routes 6-10 primarily function as feeder routes to and from U-Bahn/S-Bahn train stations.

Monte Carlo simulations are conducted to model the SoD service under each scenario, as defined by route, flexible-route portion $x_f$ (in km increment), and driver proportion (0\%, 75\%, 100\%), with 100 generated demand instances.
The simulation runs from 9~pm to midnight to capture off-peak demand, with the first hour excluded from metric evaluation to allow for warm-up.
Trip origins and destinations are derived from a boarding and alighting dataset provided by a local public transit operator.
Demand modeling follows a previously established local logit model that is sensitive to walking time \citep{moeckel_agent-based_2020}.

Parameter values used in the simulation are included in Table~\ref{tab:notation}.
Table~\ref{tab:route_data} provides an overview of route characteristics. 
Other data sources include GIS dataset of transit alignment and stop locations \citep{mvg_mvg_2024}, 
road network with steady travel time \citep{boeing_osmnx_2017}, and
SAV cost \citep{tirachini_economics_2020}.
\footnote{20-seater cost is interpolated between 8 and 44.}

    \begin{table}[thb]
    \centering
    \caption{Route characteristics and optimal service planning results}
    \label{tab:route_data}
    \begin{tabular}{|l|c|c|c|c|c|c|c|c|c|c|}
    \hline
    \textbf{Route ID} & 1 & 2 & 3 & 4 & 5 & 6 & 7 & 8 & 9 & 10 \\
    \hline
    \multicolumn{11}{|l|}{\textbf{Route Information}} \\
    \hline
    Length $L^x$ (km) & 9.9 & 5.4 & 10.7 & 11.9 & 11.5 & 6.8 & 5.6 & 4.3 & 5.8 & 3.1 \\
    Cycle Time $T^c$ (min) & 90 & 40 & 70 & 80 & 80 & 40 & 40 & 70 & 60 & 20 \\
    Peak-Hour Demand $\Lambda$ (pax/h) & 622 & 114 & 498 & 94 & 248 & 430 & 351 & 244 & 211 & 188 \\
    Off-Peak Demand (pax/h) & 227 & 65 & 77 & 20 & 134 & 110 & 138 & 25 & 36 & 59 \\
    Peak-Hour Headway $H^0$ (min) & 6 & 20 & 5 & 20 & 10 & 7.5 & 10 & 10 & 10 & 15 \\
    Vehicle size (pax) & 70 & 70 & 70 & 70 & 70 & 70 & 70 & 70 & 70 & 70 \\
    Fleet size (veh) & 15 & 2 & 14 & 4 & 8 & 6 & 4 & 7 & 6 & 2 \\
    \hline
    \multicolumn{11}{|l|}{\textbf{Optimal Results - Full SAV Scenario}} \\
    \hline
    Peak-Hour Headway $h^*$ (min) & 3 & 3.5 & 1 & 3 & 2 & 3 & 3.5 & 2 & 1.5 & 5.5 \\
    Vehicle size $b^*$ (pax) & 44 & 8 & 8 & 5 & 8 & 20 & 20 & 8 & 5 & 20 \\
    Fleet size $s^*$ (veh) & 32 & 12 & 80 & 28 & 46 & 16 & 13 & 40 & 47 & 4 \\
    Off-Peak Headway $h$ (min) & 5 & 5 & 5 & 10 & 3 & 5 & 5 & 10 & 8 & 8 \\
    \hline
    \multicolumn{11}{|l|}{\textbf{Optimal Results - 75\% Drivers ($\alpha = 0.75$)}} \\
    \hline
    Peak-Hour Headway $h^*$ (min) & 3 & 7 & 2 & 3.5 & 3.5 & 3 & 5 & 2 & 2 & 5.5 \\
    Vehicle size $b^*$ (pax) & 44 & 20 & 20 & 8 & 20 & 20 & 44 & 8 & 8 & 20 \\
    Fleet size $s^*$ (veh) & 32 & 6 & 45 & 24 & 26 & 16 & 8 & 40 & 30 & 4 \\
    Off-Peak Headway $h$ (min) & 5 & 10 & 5 & 10 & 5 & 5 & 8 & 10 & 8 & 8 \\
    \hline
    \multicolumn{11}{|l|}{\textbf{Optimal Results - 100\% Drivers ($\alpha = 1$)}} \\
    \hline
    Peak-Hour Headway $h^*$ (min) & 4 & 10 & 2.5 & 4 & 3.5 & 4 & 7 & 4 & 5 & 5.5 \\
    Vehicle size $b^*$ (pax) & 44 & 20 & 20 & 8 & 20 & 44 & 44 & 20 & 20 & 20 \\
    Fleet size $s^*$ (veh) & 25 & 4 & 34 & 20 & 25 & 10 & 6 & 18 & 13 & 4 \\
    Off-Peak Headway $h$ (min) & 5 & 15 & 5 & 10 & 5 & 6 & 10 & 10 & 8 & 10 \\
    \hline
    \end{tabular}
    \end{table}

\section{Results}
\label{sec:res}

This section begins by discussing the results of service planning for peak-hour fixed routes, considering both full SAV and transition scenarios, followed by the SoD schedule setting. 
We then analyze the simulation results of SoD service operations.
Finally, we examine the SoD use cases, deviations from theoretical values, and variations in user experience.

\subsection{Service Planning}

\subsubsection{Peak-Hour Fixed-Route Service --- Optimal Vehicle and Fleet Size (Full SAV Scenario)}

Table~\ref{tab:route_data} shows considerable improvements in service quality under the full SAV scenario.
Peak-hour headways are substantially reduced, e.g., from 6~min to 3~min for Route 1.
Besides, optimal vehicle sizes trend towards smaller.
The original 70-passenger buses are replaced by 44-passenger vehicles for Route 1, or even by smaller 8-passenger or 5-passenger cars for some other routes.
These changes, due to cost savings from SAVs, improve service quality with lower waiting times while keeping current operating costs. 
The larger fleet sizes also provide extra capacity for SoD at smaller off-peak headways, which means fewer detours in each run.

\subsubsection{Peak-Hour Fixed-Route Service --- Optimal Vehicle and Fleet Size (Transition Scenario with Drivers)}

We examined scenarios with varying percentages of drivers ($\alpha = 100\%, 75\%$)
\footnote{For 50\% drivers and lower, results converge towards the full SAV scenario as long-term capital costs are the critical factor. We also refer to Section~\ref{sec:transition} that $\alpha=1$ refers to deploying all current drivers to drive some SAVs in a bigger fleet, with some additional SAVs without drivers.}
to understand the optimal configuration during the transition.

From Table~\ref{tab:route_data}, the 100\% driver scenario results involve much higher headways and larger vehicles compared to the full SAV scenario. Instead of 8-passenger vans on Routes 2 and 3 (full SAV scenario), 20-passenger minibuses are used. 
The 75\% driver scenario falls between full drivers and SAV scenarios. 
Notably, even during the transition with drivers in SAVs, service quality improvements and waiting time reductions are achievable by switching to a fleet of smaller SAVs on less busy routes. 
Nevertheless, this is based on assumptions that vehicle sizes for a route can be switched easily (e.g., from a pool of vehicles shared by various routes). 
Otherwise, some routes may need to wait for lower driver proportions to reap the full benefits, 
which means the focus could be laid on high-demand routes first.

\subsubsection{Off-Peak Semi-on-Demand Service --- Schedule Setting}

\begin{figure}[htbp]
	\centering
	\includegraphics[width=0.6\textwidth]{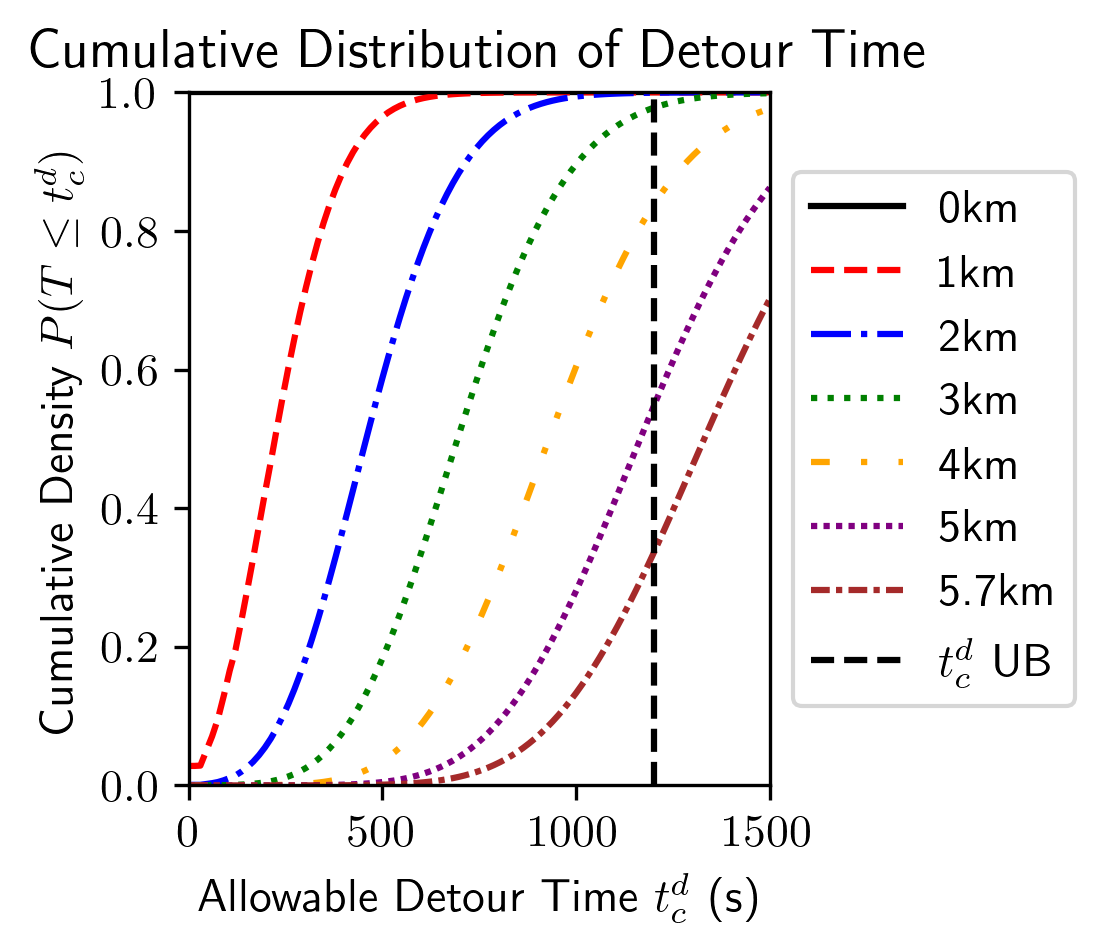}
	\caption{Cumulative distribution of detour time under varying flexible-route portions $x_f$ for Route 7 (UB refers to upper bound)}
        \label{fig:t^d_c}
    \end{figure}

Figure~\ref{fig:t^d_c} illustrates the theoretical cumulative distribution of detour time $T$ in Eq.~\eqref{eq:T_t^d_c} under different $x_f$ for Route 7 as an example.
Higher service guarantees (e.g., 95\% level) necessitate much longer vehicle detour times, which in turn require longer cycle times and more vehicles in operation.
The upper bound in Eq.~\eqref{eq:t^d_c_bound} imposed by fleet size limitations (dashed line) at longer flexible routes may lead to an unacceptable level of request rejections for too long flexible-route portions.

\subsection{Simulation of Semi-on-Demand Service}
\label{sec:sim}

\begin{figure}[htbp]
	\centering
	\includegraphics[width=0.8 \textwidth]{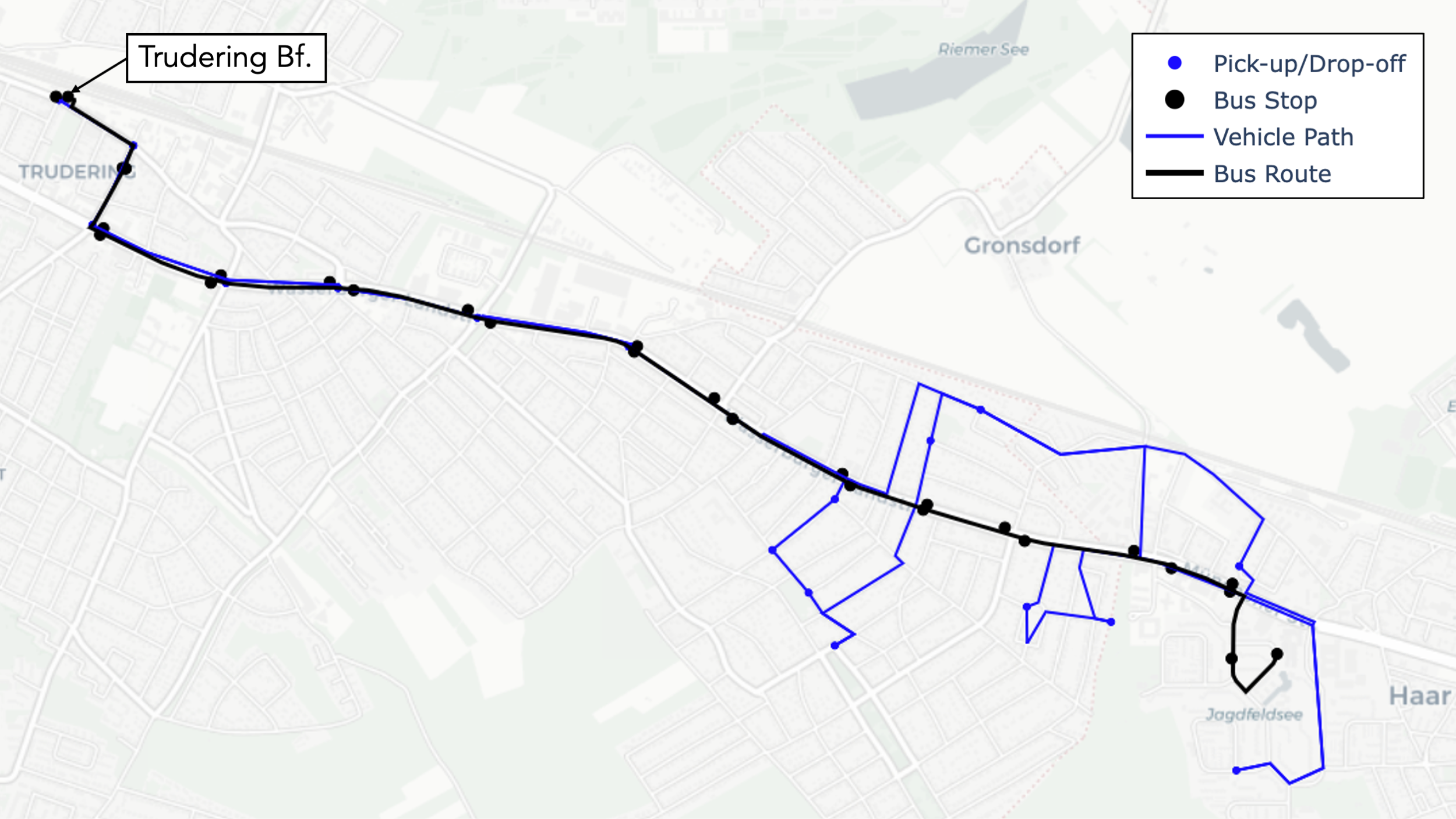}
	\caption{A simulated example of semi-on-demand hybrid route for Route 7}
        \label{fig:route_193}
    \end{figure}

A simulated SoD run of Route 7 is illustrated in Figure~\ref{fig:route_193} as an example.
After departing from the Trudering terminus and serving the fixed-route stops (black line), 
the SAV detours from the original bus route to pick up and drop off passengers in the flexible-route part (blue line).
It then returns to the fixed route and terminus.

This subsection discusses the simulation results for different routes under varying flexible-route length $x_f$.
All statistics shown are median values unless otherwise specified.

\subsubsection{Full SAV Scenario}

\begin{figure}[htbp]
    	\center
    	\includegraphics[width=\textwidth]{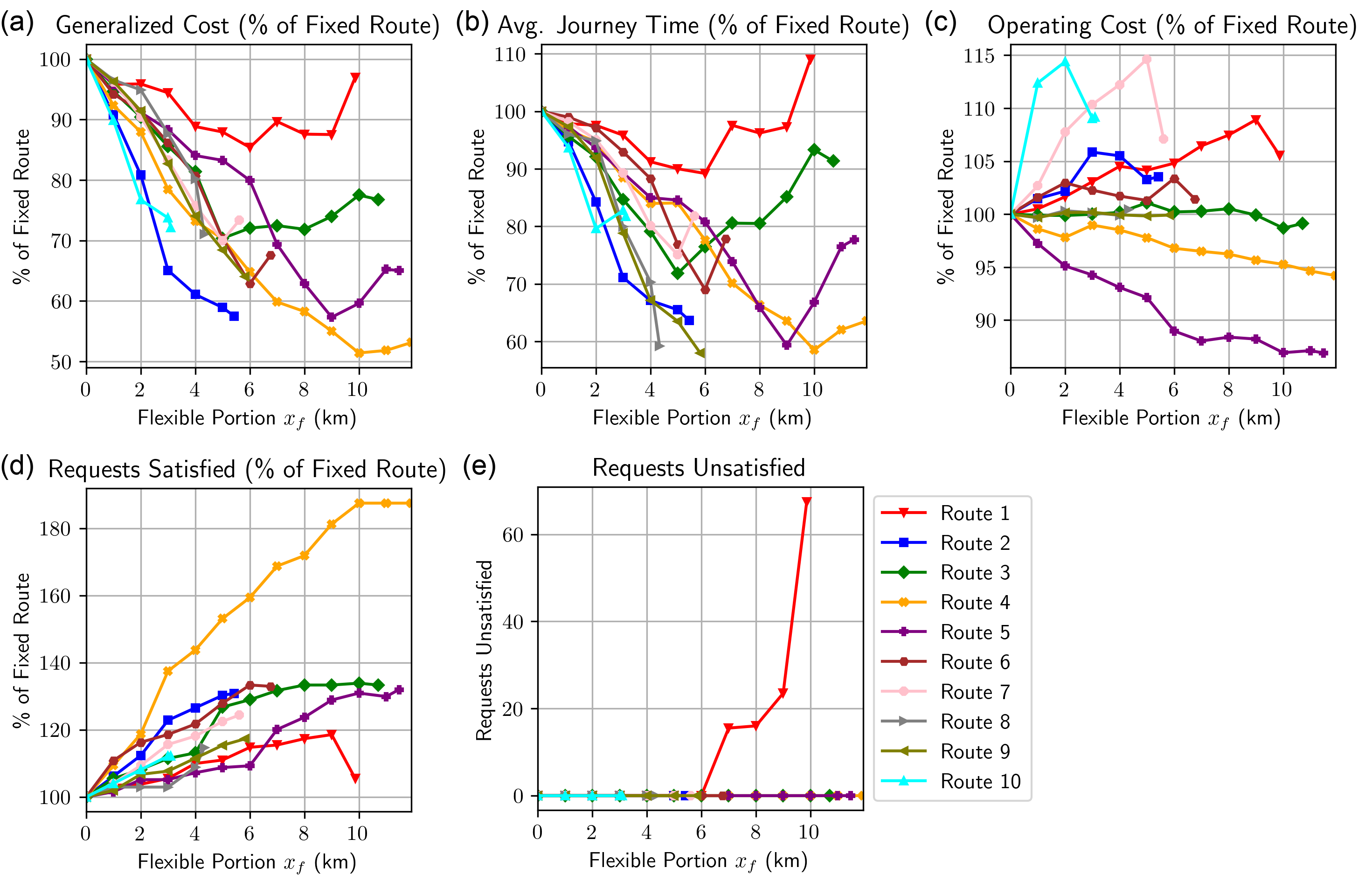}
    	\caption{Results under full SAV scenario}
            \label{fig:all_line_result}
    \end{figure}

Figure~\ref{fig:all_line_result} shows the different metrics, as a percentage of the SAV fixed-route results ($x_f=0$), under varying flexible-route portion $x_f$ across the ten routes.

The generalized costs shown in Figure~\ref{fig:all_line_result}(a) reveal that some routes (e.g., Routes 2, 8, 10) achieve optimal performance with fully flexible routing, 
while others (e.g., Routes 1, 3, 5) benefit most from limited flexible-route portions.
The savings range from 15\% (Route 1) to 50\% (Route 4).
The differences highlight the balance between detours and improvements in efficiency across different route characteristics.
These can be further analyzed from the user's and operator's perspectives.

Figure~\ref{fig:all_line_result}(b) shows the journey time changes, which are affected by the reduced access time and increased detours with longer flexible routes, as well as changes in waiting time and savings in fixed stops. 
For some routes, e.g., Routes 2 and 8, fully flexible routing brings the minimum journey times.
Other routes, like Routes 1, 3, and 5, show an initial decrease in journey time followed by a rebound.

For the operating costs in Figure~\ref{fig:all_line_result}(c), longer routes with medium to low demand, like Routes 4 and 5, show lower operating costs, 
while Routes 1, 7, 10 exhibit higher operating costs, 
likely due to higher demand and thus higher occupancy per run (all exceeding 7 passengers).

Due to improved accessibility in the flexible route, the number of requests satisfied generally increases as shown in Figure~\ref{fig:all_line_result}(d).
20\%-80\% more passengers are served.
However, for Route 1, this increase is counterbalanced by excessive detours at longer $x_f$ (likely related to its higher ridership).
As seen in Figure~\ref{fig:all_line_result}(e), up to 65 requests cannot be served when $x_f$ exceeds 6km for this route.
Meanwhile, for Route 4, flexible-route service gets rid of the circuitous loop, significantly improving efficiency and accessibility (see Figure~\ref{fig:exp_buslines}).

Detailed results are presented in Table~\ref{tab:result} ($\alpha=0$) after References.

\subsubsection{Transition Scenarios with Drivers}

\begin{figure}[htbp]
    	\center
    	\includegraphics[width=\textwidth]{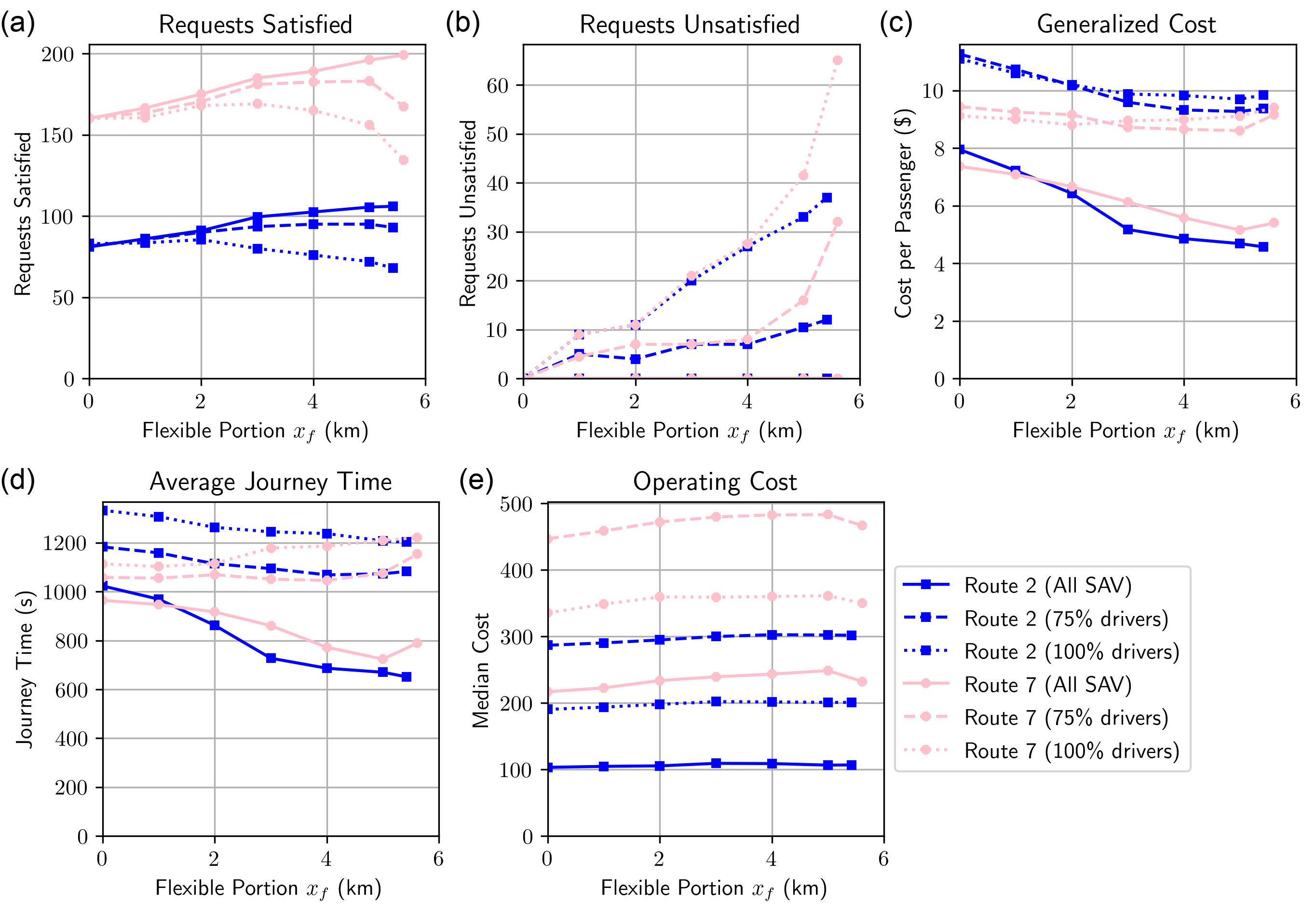}
    	\caption{Results under scenarios of full SAV, transition with 75\% and 100\% drivers}
            \label{fig:two_line}
    \end{figure}

Figure~\ref{fig:two_line} shows only Routes 2 and 7 for clarity, illustrating the differences in the transition and full SAV stage.
\footnote{The detailed results for all routes are shown as $\alpha=0.75$ and $\alpha=1$ cases in Table~\ref{tab:result} after References.}
Figure~\ref{fig:two_line}(a) shows that the full SAV case satisfies more requests with longer flexible routes due to improved accessibility.
However, for the transition scenarios with 75\% and 100\% drivers, requests satisfied plateau at respectively 5km and 2km flexible-route length. 
This is attributed to lower headways in the transition scenarios (Table~\ref{tab:route_data}),
resulting in more passengers per run and consequently more detours. 
These factors limit the operational capacity of flexible routes, leading to more request rejection.

Generalized costs, depicted in Figure~\ref{fig:two_line}(c), show that Route 2 still benefits from moderate flexible routing in the transition stage, 
while the benefits of Route 7 are less visible. 
In Figure~\ref{fig:two_line}(d), the scenarios with more drivers feature higher journey times due to increased waiting times and detours.
Besides, these scenarios have higher operating costs due to the labor costs (Figure~\ref{fig:two_line}(e)).

Overall, the transition scenarios with drivers limit the SAV benefits in terms of generalized cost reduction and additional passengers served.
Nevertheless, adding SAVs parallel to drivers and implementing SoD still offer advantages over existing fixed routes, including increased demand and improved accessibility, particularly at shorter flexible-route portions. 
The hybrid-route approach provides the flexibility to gradually adjust the flexible-route portion in response to changes in the driver proportion in the fleet.

\subsection{Analysis}

\subsubsection{Use Cases}
\begin{figure}[htbp]
    	\center
    	\includegraphics[width=0.5\textwidth]{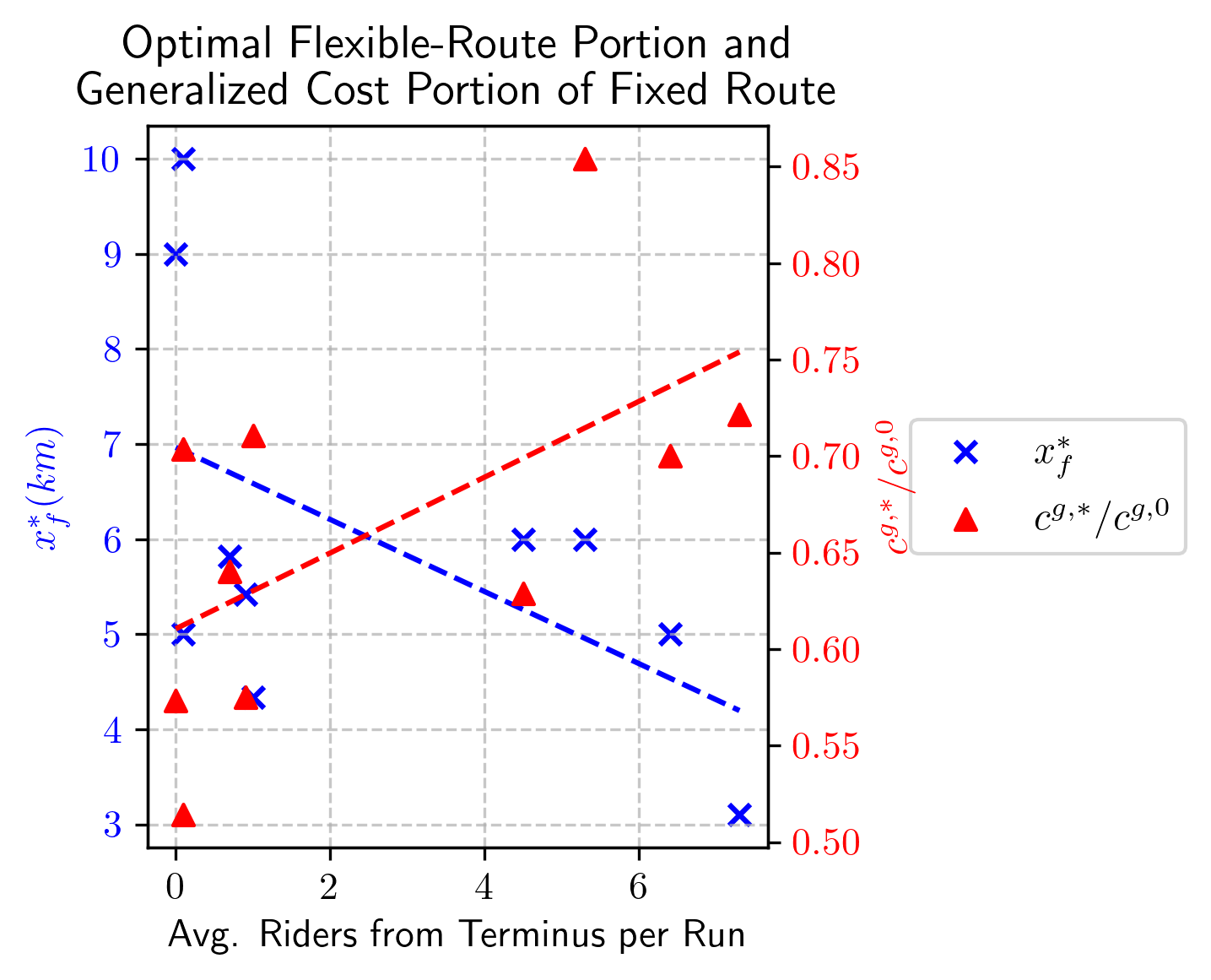}
    	\caption{Optimal flexible-route portion and generalized cost portion of fixed route}
            \label{fig:pax_term_run}
    \end{figure}

The implementation and benefits of SoD routes vary significantly across different routes due to specific factors such as demand density and distribution, operating headway, and route length. 
A clearer trend emerges when we examine the number of riders departing from the terminus per run, as illustrated in Figure~\ref{fig:pax_term_run}.

The figure shows that routes with higher ridership from the terminus per run tend to have shorter optimal flexible-route portions $x_f^*$ and smaller reductions in generalized cost (higher $c^{g,*}/c^{g,0}$).
This pattern may be explained by the better scale of economy of fixed routes under more concentrated demand. 
Additionally, routes with more passengers per run experience more detours in excessively long flexible routes.

Given the complexity and variability of these factors, it is advisable to utilize the analytical formula \citep{ng_semi--demand_2024} and run simulations for each route under various implementation scenarios to determine the optimal strategy.

\subsubsection{Comparison with Theoretical Results}

\begin{figure}[htbp]
	\center
	\includegraphics[width=0.5 \textwidth]{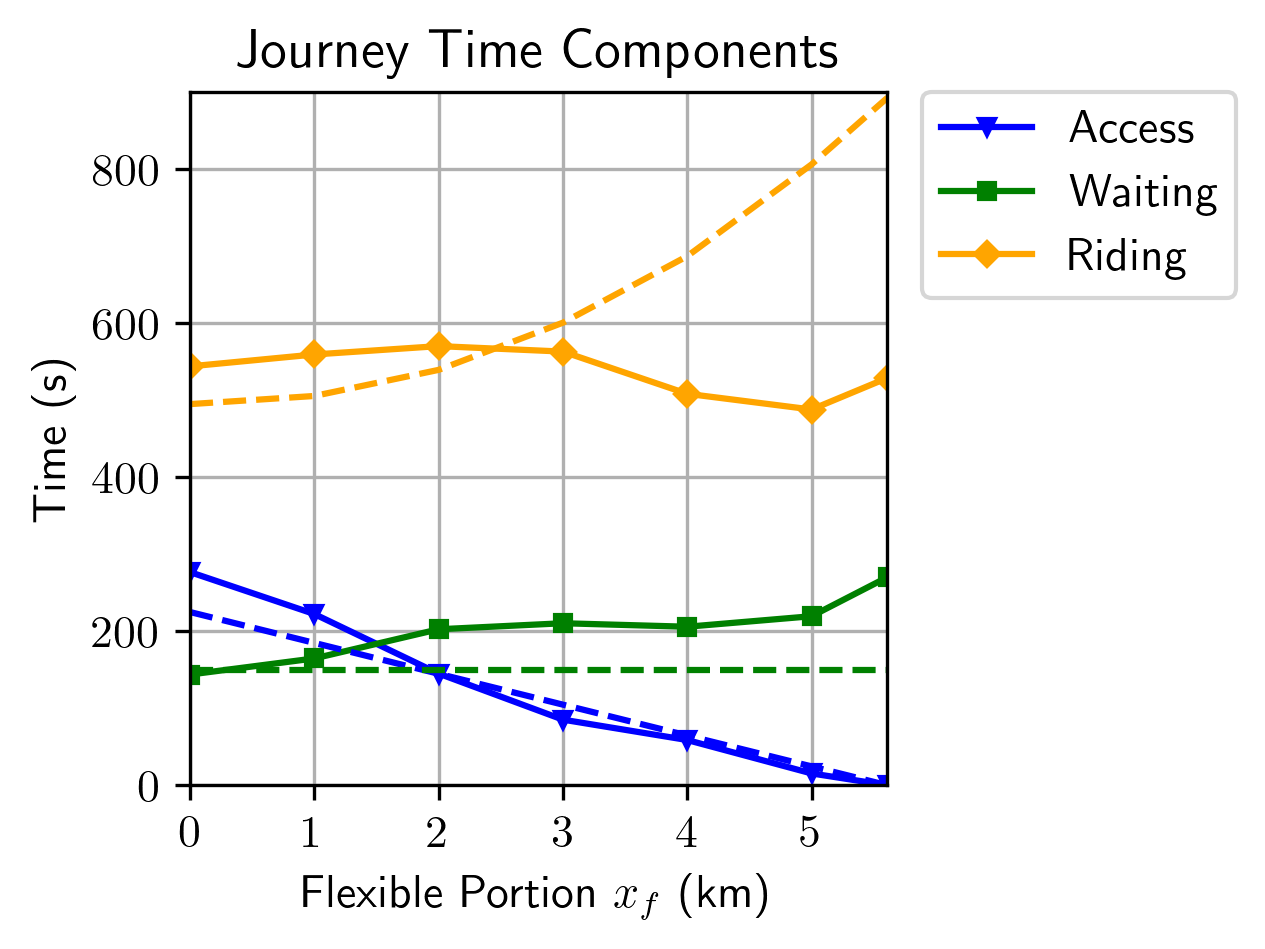}
	\caption{Comparison between results of simulation (solid lines) and theoretical (dashed lines) values of Route 7 (full SAV scenario)}
        \label{fig:193_result}
\end{figure}

Figure~\ref{fig:193_result} illustrates the trade-offs among various passenger time metrics (solid lines) as the flexible-route portion increases, 
compared with theoretical values assuming a uniform demand distribution (dashed lines) for Route 7.

Access (walking) times decrease with longer flexible routes, indicating improved service accessibility that aligns with theoretical predictions.
Waiting times increase gradually with $x_f$ due to the effects of detours on headway variance. 
Riding times fluctuate around the same level, contrary to the theoretical quadratic increase, possibly reflecting the limit set by the constraint of the detour factor.

\subsubsection{User Experience Analysis}

\begin{figure}[htbp]
    	\center
    	\includegraphics[width=\textwidth]{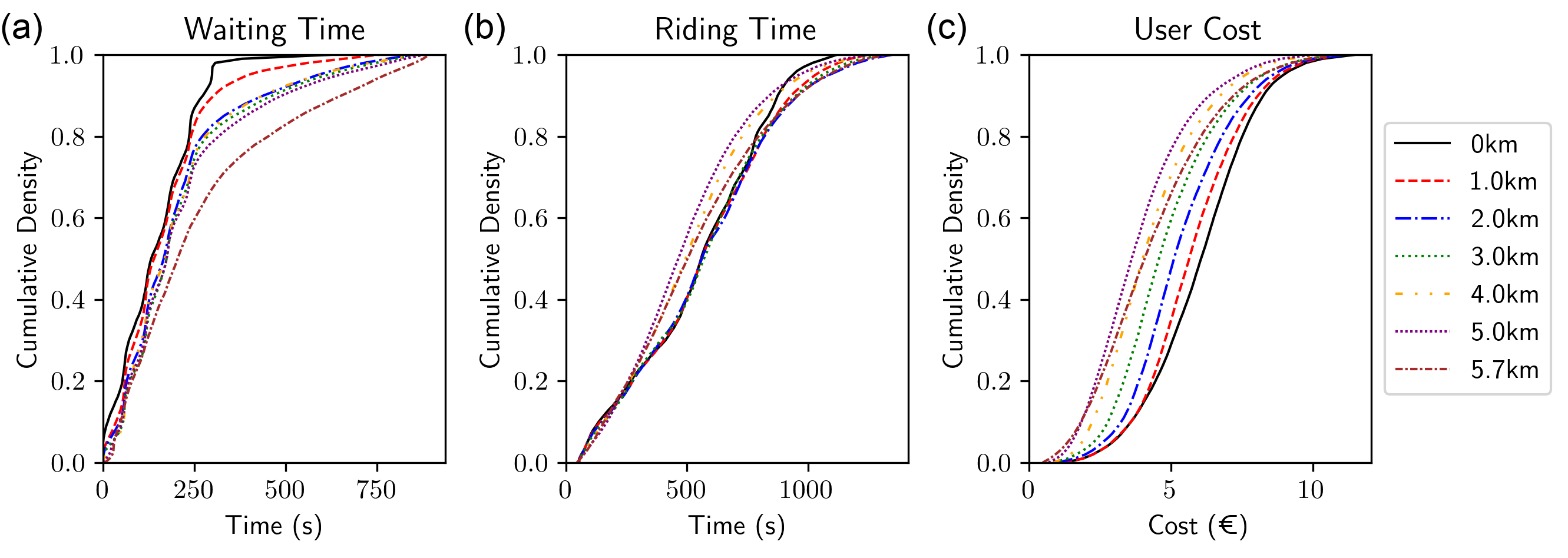}
    	\caption{Distribution of user experience under varying flexible-route portions $x_f$ of Route 7 (full SAV scenario)}
            \label{fig:193_user}
    \end{figure}

Would SoD routes bring different user experiences for different riders?
Figure~\ref{fig:193_user} shows the cumulative distribution of user experiences under varying flexible-route portions $x_f$.

Waiting time changes (Figure~\ref{fig:193_user}(a)) reveal that the fixed route (0km) offers the lowest and most consistent waiting times. 
As the flexible-route portion increases, waiting times gradually rise, with the higher end experiencing considerably longer waiting times (approaching the 15-minute limit).
Fully flexible routes perform the worst in terms of waiting time consistency.

In Figure~\ref{fig:193_user}(b), riding time decreases slightly with longer flexible routes, likely due to savings in fixed dwell times.
However, this effect is relatively modest and evenly distributed across users.
Examining overall user cost (access + waiting + riding) in Figure~\ref{fig:193_user}(c), we observe a seemingly balanced outcome. 
Users who benefit from reduced walking distances often experience longer waiting times, particularly at the flexible end of the route. 
Despite this trade-off, the overall impact appears beneficial for most users.

This example showcases the use of cumulative distribution charts to analyze the variation in user experience implementing SoD service in each route.
Other equity metrics \citep{mahmassani_improving_2024} can provide a more comprehensive picture of its impacts during planning.

\section{Conclusion}
\label{sec:concl}

This study addresses research gaps in semi-on-demand (SoD) service planning, simulation, and analysis with shared autonomous vehicles (SAVs). 
We focused on improving current transit services by optimizing peak-hour fixed-route operations and implementing off-peak SoD services in the same route catchment area.
Our research addresses four key aspects:
(i) determining optimal fleet and vehicle sizes for peak-hour fixed-route services with SAVs,
(ii) assessing the impact of operating budgets during the transition from human drivers to full automation,
(iii) developing optimal service planning for off-peak SoD hybrid-route services in schedule and flexible-route length, and 
(iv) identifying suitable use cases for off-peak SoD services. 

For peak-hour fixed routes, we analytically optimize fleet and vehicle sizes that minimize users' waiting time and operator's cost under budget and capacity constraints.
Off-peak SoD service is conceptualized with analytical schedule settings and analyzed with an agent-based simulation on ten existing bus routes in Munich, Germany. 
Results corroborate the SoD benefits, satisfying up to 80\% more requests and saving up to 50\% generalized cost when using SAVs. 
At moderate flexible-route lengths, SoD routes avoid excessive detours and attract more demand with the door-to-door convenience.
Cross-route comparison and user experience analysis enhance the precision of use case delineation of SoD with SAVs.

Our findings highlight that SAVs can bring considerable benefits in fixed-route service quality through increased frequency and smaller vehicle sizes, even during the transition with a partial or full human driver workforce.
The increased fleet size and reduced headways favor more flexible-route operation in the SoD service.
In the case study, some routes benefit from fully flexible operations, 
while others still require some fixed-route portion to avoid excessive detours and operational burden to serve all passengers.
During the transition phase, the limited fleet size and higher headway constrain the benefits of flexible-route operations.
Nevertheless, SoD hybrid routes offer a good transition strategy,
allowing for gradual adjustment of the flexible-route portion when more SAVs are deployed.
Notably, for all routes studied, optimal fleet and vehicle sizes with 50\% or more SAVs converge to the full SAV solution, 
indicating that SAV capital cost, rather than day-to-day operating costs, becomes the decision factor in the later stages of transition.
Furthermore, our analysis identifies that average occupancy per run from the terminus, among other demand and operational characteristics, is an important factor in determining the benefits of SoD.

We also highlight several limitations in this study for further research:
    \begin{itemize}
    \item Cross-line effects, including demand and fleet assignment across multiple routes, call for more holistic optimization.
    \item The interaction between SoD services and ride-pooling/ride-sharing systems may offer synergy or competition.
    \item Further optimization of headway settings could enhance the efficiency of SoD operations.
    \item The feasibility of larger SAVs navigating residential areas for pick-up and drop-off needs further investigation, although smaller vehicles are expected with full SAV deployment.
    \end{itemize}

In conclusion, this study demonstrates the potential of SoD hybrid-route services using SAVs to enhance public transit systems with higher efficiency and better accessibility. 
The step-by-step service planning, operational simulation, and analysis framework, coupled with previous theoretical approach \citep{ng_semi--demand_2024}, supports transit agencies in planning and evaluating the strategy to roll out SoD services with SAVs in existing networks.

\section*{Acknowledgements}
A preliminary version of this work was presented at the Conference in Emerging Technologies in Transportation Systems (TRC-30). 
The authors are grateful for the valuable comments from the editors and reviewers that have helped to improve this paper.
This work is based on research funded in part by the German Academic Exchange Service and Northwestern Buffett Institute for Global Affairs to the first author. 
Additionally, this work is partially funded by the German Federal Ministry of Education and Research via the MCube Project STEAM.

\bibliographystyle{elsarticle-harv} 
\bibliography{references}

\section*{Simulation Result Table}
Table~\ref{tab:result} shows the simulation results discussed in Section~\ref{sec:sim}.
All metrics shown are median values. 
$\alpha=0.75$ case is not shown if equivalent to $\alpha=0$.

\begin{landscape}
\begin{longtable}{|c|c|c|c|c|cccc|c|c|}
\caption{Simulation Results}
\label{tab:result}\\
\hline
\multirow{2}{*}{Route} & \multirow{2}{*}{$\alpha$} & \multirow{2}{*}{\begin{tabular}[c]{@{}c@{}}$x_f$\\ (km)\end{tabular}} & \multirow{2}{*}{Requests} & \multirow{2}{*}{\begin{tabular}[c]{@{}c@{}}Requests\\ satisfied\end{tabular}} & \multicolumn{4}{c|}{Avg. Time (s)}                                                            & \multirow{2}{*}{\begin{tabular}[c]{@{}c@{}}Veh.\\ dist. (km)\end{tabular}} & \multirow{2}{*}{\begin{tabular}[c]{@{}c@{}}Avg. gen.\\ cost (€)\end{tabular}} \\ \cline{6-9}
                       &                           &                                                                       &                           &                                                                               & \multicolumn{1}{c|}{Ride}  & \multicolumn{1}{c|}{Wait}  & \multicolumn{1}{c|}{Walk} & Journey &                                                                            &                                                                               \\ \hline
\endfirsthead
\multicolumn{11}{c}%
{{\bfseries Table \thetable\ continued from previous page}} \\
\hline
\multirow{2}{*}{Route} & \multirow{2}{*}{$\alpha$} & \multirow{2}{*}{\begin{tabular}[c]{@{}c@{}}$x_f$\\ (km)\end{tabular}} & \multirow{2}{*}{Requests} & \multirow{2}{*}{\begin{tabular}[c]{@{}c@{}}Requests\\ satisfied\end{tabular}} & \multicolumn{4}{c|}{Avg. Time (s)}                                                            & \multirow{2}{*}{\begin{tabular}[c]{@{}c@{}}Veh.\\ dist. (km)\end{tabular}} & \multirow{2}{*}{\begin{tabular}[c]{@{}c@{}}Avg. gen.\\ cost (€)\end{tabular}} \\ \cline{6-9}
                       &                           &                                                                       &                           &                                                                               & \multicolumn{1}{c|}{Ride}  & \multicolumn{1}{c|}{Wait}  & \multicolumn{1}{c|}{Walk} & Journey &                                                                            &                                                                               \\ \hline
\endhead
1                                            & 0                                              & 0.0                                                                                        & 291                                            & 291                                                                                                & \multicolumn{1}{c|}{10.21} & \multicolumn{1}{c|}{2.44}  & \multicolumn{1}{c|}{5.21} & 17.89                        & 773.1                                                                                           & 9.35                                                                                               \\ \hline
1                                            & 0                                              & 1.0                                                                                        & 301                                            & 301                                                                                                & \multicolumn{1}{c|}{10.53} & \multicolumn{1}{c|}{2.55}  & \multicolumn{1}{c|}{4.42} & 17.51                        & 777.0                                                                                           & 8.96                                                                                               \\ \hline
1                                            & 0                                              & 2.0                                                                                        & 302                                            & 302                                                                                                & \multicolumn{1}{c|}{10.32} & \multicolumn{1}{c|}{2.82}  & \multicolumn{1}{c|}{4.27} & 17.46                        & 785.8                                                                                           & 8.97                                                                                               \\
\hline
\end{longtable}
\end{landscape}

\end{document}